%% file: LDCodec.tex
\documentclass[conference]{IEEEtran}
\IEEEoverridecommandlockouts
\usepackage{cite}
\usepackage{amsmath,amssymb,amsfonts, subcaption,url,booktabs}
\usepackage{algorithmic}
\usepackage{graphicx}
\usepackage{textcomp}
\usepackage{xcolor}
\usepackage{tabularx,booktabs}
\usepackage{multirow}
\usepackage{array}

\def\BibTeX{{\rm B\kern-.05em{\sc i\kern-.025em b}\kern-.08em
    T\kern-.1667em\lower.7ex\hbox{E}\kern-.125emX}}

\makeatletter
\newcommand{\linebreakand}{%
  \end{@IEEEauthorhalign}
  \hfill\mbox{}\par
  \mbox{}\hfill\begin{@IEEEauthorhalign}
}
\makeatother

\title{LDCodec: A high quality neural audio codec with low-complexity decoder}

\author{
\IEEEauthorblockN{1\textsuperscript{st} Jiawei Jiang}
\IEEEauthorblockA{\textit{ByteDance China}\\
Beijing, China \\
jiangjiawei.lahm@bytedance.com}
\and
\IEEEauthorblockN{2\textsuperscript{nd} Linping Xu}
\IEEEauthorblockA{\textit{ByteDance China}\\
Beijing, China \\
xulinping.678@bytedance.com}
\and
\IEEEauthorblockN{3\textsuperscript{rd} Dejun Zhang}
\IEEEauthorblockA{\textit{ByteDance China}\\
Beijing, China \\
zhangdejun@bytedance.com}
\linebreakand
\IEEEauthorblockN{4\textsuperscript{th} Qingbo Huang}
\IEEEauthorblockA{\textit{ByteDance China}\\
Beijing, China \\
qingbohuang@bytedance.com}
\and
\IEEEauthorblockN{5\textsuperscript{th} Xianjun Xia}
\IEEEauthorblockA{\textit{ByteDance China}\\
Shenzhen, China \\
xiaxianjun@bytedance.com}
\and
\IEEEauthorblockN{6\textsuperscript{th} Yijian Xiao}
\IEEEauthorblockA{\textit{ByteDance China}\\
Shenzhen, China \\
xiaoyijian@bytedance.com}
}

\begin{document}
\maketitle

\begin{abstract}
Neural audio coding has been shown to outperform classical audio coding at extremely low bitrates. However, the practical application of neural audio codecs is still limited by their elevated complexity. To address this challenge, we have developed a high-quality neural audio codec with a low-complexity decoder, named LDCodec (Low-complexity Decoder Neural Audio Codec), specifically designed for on-demand streaming media clients, such as smartphones. 
Specifically, we introduced a novel residual unit combined with Long-term and Short-term Residual Vector Quantization (LSRVQ), subband-fullband frequency discriminators, and perceptual loss functions. This combination results in high-quality audio reconstruction with lower complexity. 
Both our subjective and objective tests demonstrated that our proposed LDCodec at 6kbps outperforms Opus at 12kbps.
\end{abstract}


\section{Introduction}
\label{LDCodec_intro}
\input{sections/introduction}

\begin{figure*}[h]
  \vspace{0.2cm}
  \begin{minipage}[b]{1.0\linewidth}
    \centering
    \includegraphics[width=1\textwidth]{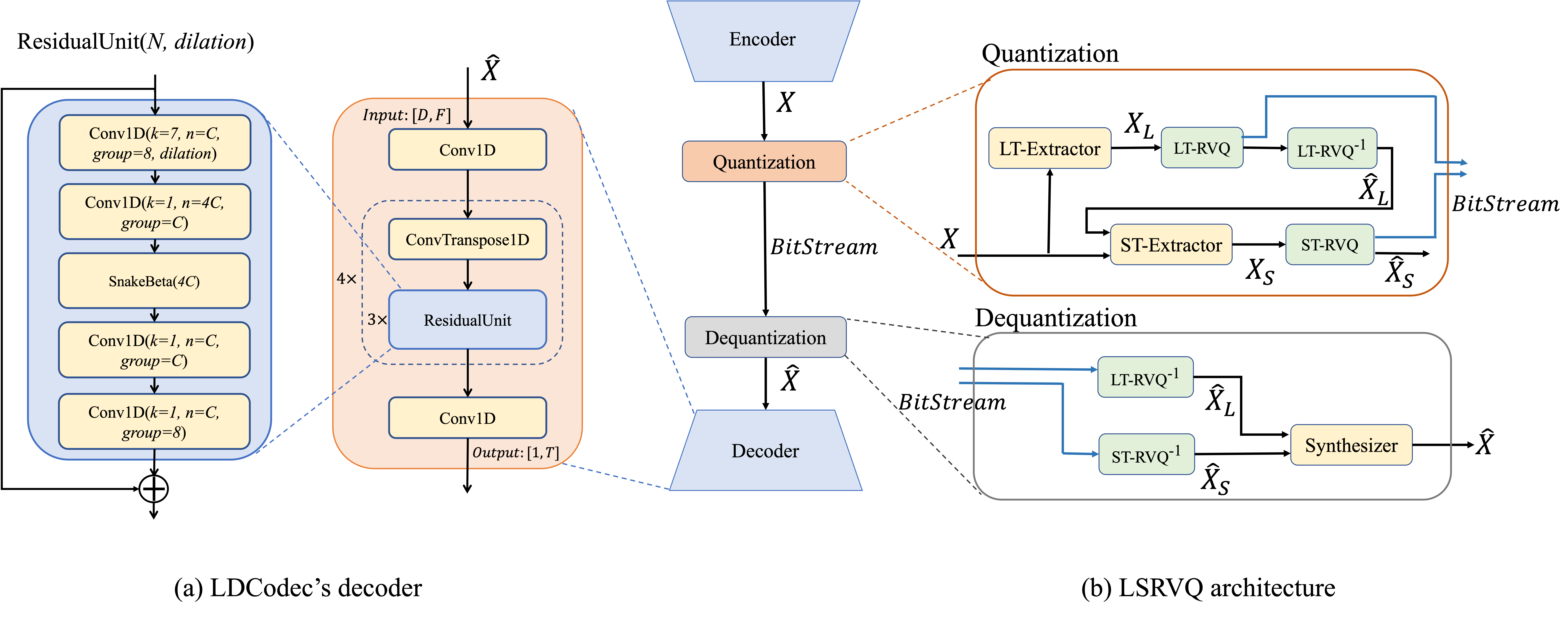}
  \end{minipage}
  \caption{
    Illustration of the proposed model. (a) The decoder structure of LDCodec. (b) The quantizer architecture (LSRVQ).
  }
  \label{fig:LDCodec_structure}
\end{figure*}
\section{Proposed Audio Codec}
\label{LDCodec_codec}
\input{sections/architecture}

\section{Experiments}
\input{sections/experiments}

\section{Conclusion}
\label{LDCodec_conclusion}

To make neural audio codecs applicable on devices with limited computational resources, we introduced LDCodec, a high-quality neural audio codec with low decoding complexity. Our innovation lied in the design of the residual unit that enables high-quality audio reconstruction with minimal complexity cost. We also incorporated LSRVQ, subband-fullband frequency discriminators and perceptual loss functions to enhance coding performance.
Future work will focus on improving coding quality and further reducing computational complexity for streaming media clients.
\newpage

\bibliographystyle{IEEEtran}
\bibliography{LDCodec}

\end{document}

%% file: sections/introduction.tex
Audio codec technologies are fundamental to on-demand streaming media scenarios. They enable streaming media companies to distribute a wide variety of high quality audio content to their users with minimal storage and bandwidth cost. Considering the audio decoding process is typically carried out on mobile devices, an audio codec that offers high fidelity, outstanding coding efficiency, and low decoding complexity is crucial for ensuring an optimal user experience in streaming media services.

End-to-end neural audio codecs with learnable encoders, such as Soundstream~\cite{soundstream}, Encodec~\cite{defossez2022high}, and DAC~\cite{kumar2024high}, have drawn substantial interest from the research community due to their capability of delivering high-quality audio at extremely low bitrates, which is hard to achieve with traditional methods.
Soundstream utilizes a neural network-based encoder and decoder framework and is capable of encoding audio at bitrates between 3 kbps and 12 kbps due to the structured dropout applied to RVQ training.
Encodec follows the Soundstream recipe and improves audio quality by introducing a multiscale STFT discriminator and a multiscale spectral reconstruction loss. 
DAC incorporates periodic inductive biases, enhancing codebook learning through low-dimensional space projections, and introducing a multi-scale subband STFT discriminator.

Despite the high-quality generating ability, Soundstream, Encodec, and DAC share significant drawbacks of large number of parameters and high computational cost, which reaches several Giga Multiply-Add Operations per Second (GMACs). Additionally, their performance declines sharply when model complexity is reduced, making them less applicable on devices with limited computational resources, such as smartphones. To address this issue, alternatives like Lyra2\footnote{https://github.com/google/lyra}, FunCodec~\cite{du2024funcodec}, and LightCodec~\cite{xu2024lightcodec} have emerged.
Lyra2 minimizes computational complexity by employing group convolution and replacing the final upsampling decoder unit with a simple convolution layer. FunCodec uses depthwise convolutions to reduce both parameter number and computational complexity. LightCodec takes a different approach by utilizing frequency band division and a unique structure called WBCBI to reduce model complexity, while a compensation module corrects quantization errors. However, these solutions still fall short of achieving a satisfactory quality when faced with low complexity requirements.

To achieve high-quality audio reconstruction while minimizing computational cost, we proposed a low-complexity but high quality neural audio codec, LDCodec. We introduced four strategies: 1) designing an innovative residual unit composed of an expanding layer, SnakeBeta activation layer and a shrinking layer; 2) introducing LSRVQ which quantizes long-term and short-term features to utilize inter-frame correlations; 3) introducing subband-fullband frequency discriminators to reduce encoding quantization error in high-frequency; and 4) utilizing perceptual loss functions to improve transient modeling~\cite{edler2006detection} and penalize excess reconstructed energy. 
We compared qualities of different audio codecs, and our experiments demonstrated the effectiveness of our proposed methods. Remarkably, LDCodec shows a low complexity of only 0.26 GMACs in decoding process. 

%% file: sections/architecture.tex
In this section, an overview of our proposed LDCodec is introduced and the details of each module are also described. Our model follows the mainstream of end-to-end neural codecs and contains an encoder, a quantizer and a decoder shown in Fig \ref{fig:LDCodec_structure}. Section 2.1 elaborates on the encoder and a precisely designed decoder in our LDCodec. Section 2.2 introduces the long-term and short-term residual vector quantization, denoted as LSRVQ. Section 2.3 details the subband-fullband discriminators for adversarial training. Section 2.4 presents the frequency domain perceptual loss, which is utilized to improve transient modeling and penalize excess reconstructed energy.

The encoder takes the audio waveform $\in \mathbb{R}^{T}$ as input and transforms it into a feature $X \in \mathbb{R}^{D \times F}$, where  $T$ represents the audio duration, $D$ represents the feature dimension, and $F$ represents the frame number. The quantizer then converts the feature $X$ into the quantized latent feature $\hat{X}$. The decoder finally translates $\hat{X}$ back to the audio waveform.

\subsection{Encoder and Decoder}

Our encoder aligns with DAC's~\cite{kumar2024high} architecture. Figure \ref{fig:LDCodec_structure}(a) shows the decoder model structure of our proposed LDCodec.
We utilize four decoder blocks, and the upsampling factor $r$ for the ConvTranspose layer is set at $\{$8, 5, 4, 2$\}$. Each decoder block halves the number of channels and incorporates three dilated residual units with a dilation at $\{$1, 3, 9$\}$. Finally, we apply a convolution layer with a tanh activation function to convert the feature back to the audio signal. We employ group convolutions in the decoder to reduce computational costs.

Our proposed residual units incorporate two special designs. Firstly, we substitute the DAC's Snake activation with SnakeBeta to enhance periodic components.

\begin{equation}
  \mathrm{SnakeBeta}(x)=x+\frac{1}{\beta } \sin ^{2} (\alpha x)
\end{equation}

Secondly, inspired by the BigVGAN~\cite{lee2022bigvgan}, we employ a Conv1D layer to expand the latent feature in order to obtain more periodic components, and also use another Conv1D layer to shrink the feature processed by SnakeBeta activation. In our experiments, we observe such design significantly enhances audio quality compared with simply reducing the decoder-channel in DAC.

\begin{figure}[t]
  \begin{minipage}[b]{1.0\linewidth}
    \centering
    \centerline{\includegraphics[width=1\linewidth]{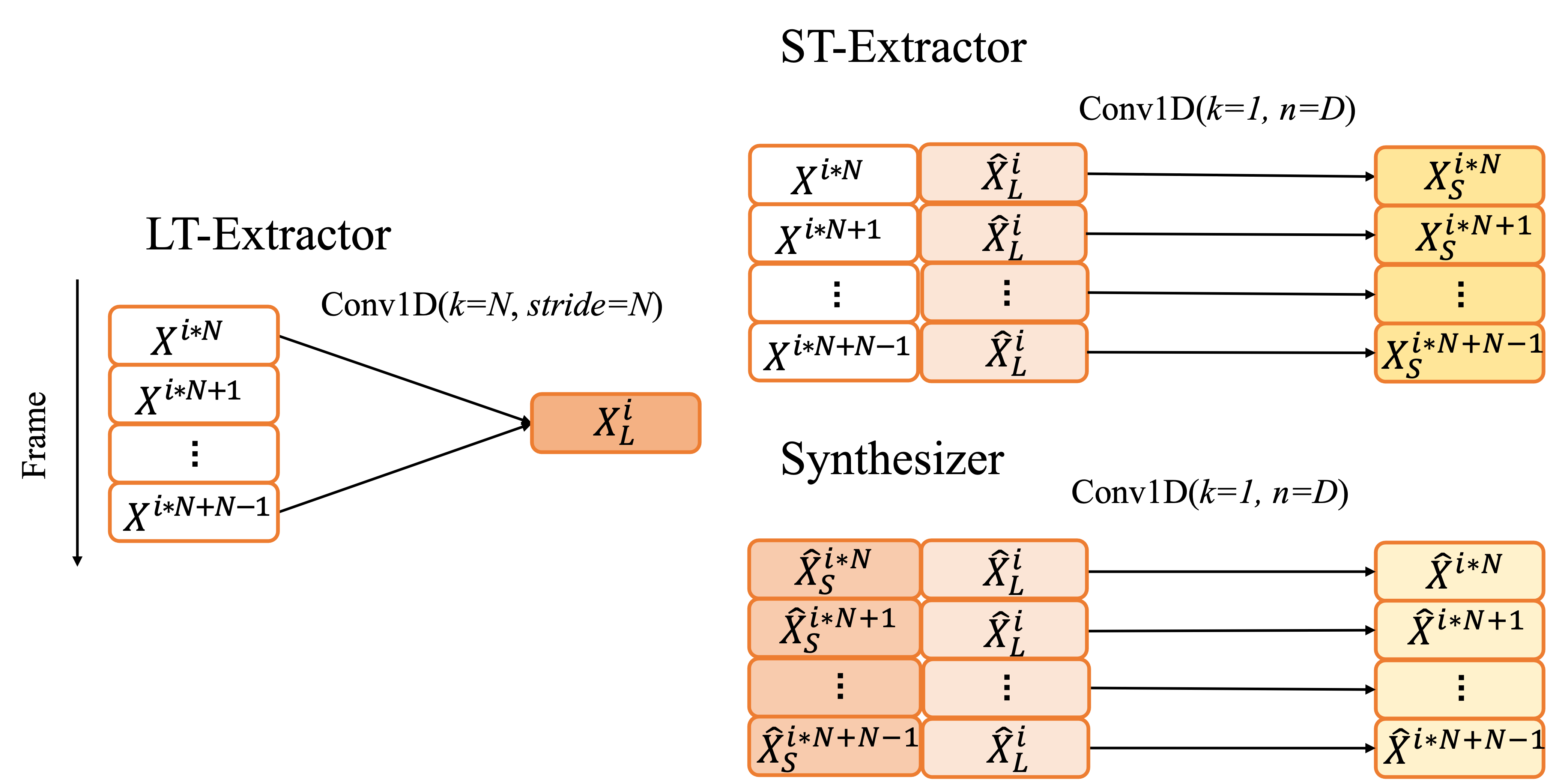}}
  \end{minipage}
  \caption{
    Detail of the Conv1D based feature LT-Extractor, ST-Extractor and Synthesizer in LSRVQ.
  }
  \label{LSRVQ_detail}
  \end{figure}

\subsection{Quantizer: LSRVQ}
Among prominent codecs like Soundstream and DAC, the quantizer compresses each audio feature individually, overlooking the temporal correlations between inter-frame features. Various methods seek to leverage these correlations to boost quantizer efficiency.
Predictive TF-Codec~\cite{jiang2023latent} proposes latent-domain predictive coding to remove temporal redundancies. TiCodec~\cite{ren2024fewer} extracts the time-invariant information from an utterance and quantizes it into a separate code to avoid repetitive information transmission. SNAC~\cite{snac} introduces multi-scale RVQ, uses coarser tokens with a wider time span and a lower sampling rate to reduce bitrate.

Motivated by those works, we introduce a unique multi-scale quantizer LSRVQ in this paper. This quantizer separates audio features into long-term and short-term groups and then quantizes them individually using LT-RVQ and ST-RVQ. The structure of LSRVQ is illustrated in Figure \ref{fig:LDCodec_structure}(b).

During the encoding process, The LT-Extractor gets the long-term feature $X_{L}$ from multi-frame features with the quantization step $N$, given by $X_{L}=f_{\mathrm{LT\text{-}Extractor} }(X)$ and $X_{L}$ is quantized by LT-RVQ. ST-Extractor extracts the residual short-term information $X_{S}$, given by $X_{S}=f_{\mathrm{ST\text{-}Extractor} }(X,\hat{X_{L}})$, then $X_{S}$ is quantized by ST-RVQ.
During the decoding process, Synthesizer merges quantized long-term feature $\hat{X_{L}}$ with quantized short-term feature $\hat{X_{S}}$ to generate the reconstructed latent feature $\hat{X}$, given by $\hat{X}=f_{\mathrm{Synthesizer} }(\hat{X_{L}},\hat{X_{S}})$.

LT-RVQ cascades $M_{q1}$ layers of VQ and each quantizer uses a codebook size of $M_{1}$, while ST-RVQ uses $M_{q2}$ layers and each codebook size is $M_{2}$. The target bitrate $B$ is:

\vspace{-0.4cm}
\begin{equation}
  B=\frac{S}{N}*M_{q1}*\log_{2}{M_{1}} + S*M_{q2}*\log_{2}{M_{2}}
\end{equation}

where $S$ denotes the input frame rate.

Figure \ref{LSRVQ_detail} shows the schematic diagrams of the Conv1D based LT-Extractor, ST-Extractor, and Synthesizer. The AvgPooling based LT-Extractor and ST-Extractor are also explored in ablation experiments.

In addition, we apply the Beam-search algorithm~\cite{xu23_interspeech} to LSRVQ, which improves the quantization efficiency without adding complexity to the decoding process.

\subsection{Discriminator}
We employ multi-period waveform discriminators (MPD)~\cite{kong2020hifi} and multi-resolution spectrogram discriminators (MRSD)~\cite{jang2021univnet} to improve audio fidelity. DAC~\cite{kumar2024high} demonstrated that using complex STFT discriminators enhances phase modeling and splitting the STFT spectrogram into sub-bands improves high frequency prediction.

We discover that splitting the STFT spectrogram into sub-band enables different convolution kernels to learn varied weight parameters and emphasize different patterns in different frequency bands. However, this sometimes causes frequency pattern mismatch across neighboring subbands.

To tackle this problem, multi-resolution subband-fullband frequency discriminators are introduced. Each frequency discriminator begins by segmenting the STFT spectrogram into several bands, with each subband modeled by its own distinct convolution layers. These features are then combined into the full band and additional convolution layers are used to model the overall spectrogram. This approach allows for the detailed analysis of the audio signals within individual subbands while also ensuring a cohesive representation of the full audio signal.

\subsection{Loss function}
The multi-scale mel-reconstruction spectral loss \cite{yamamoto2020parallel} has been recognized for enhancing stability, fidelity, and convergence speed. In our model, we implement two additional strategies to further refine the spectral loss.

Initially, we observe that the reconstruction of the transient signal~\cite{edler2006detection} remains subpar. To address this, we apply a transient detection algorithm and categorize each audio frame as either a transient or non-transient. When computing frequency reconstruction loss, we prioritize the recovery terms of transient energy to improve transient modeling.
\begin{gather}
  \mathcal L_{\mathrm{mel\text{-}transient} }=E_{(s,t,f)} [\left \|\lambda_{t}(\phi_{t,f}^{s} (x)-\phi_{t,f}^{s}(G(x))  )\right \|_{1}  ] \\
\lambda_{t}=\left\{\begin{matrix}
2,  & x\,\in \,transient\\
1, & x\,\notin \,transient
\end{matrix}\right.
\end{gather}

where $\phi_{t,f}^{s}$ denotes the $t$-th frame and $f$-th frequency bin computed from the $s$-th multi-resolution function that converts the waveform into the log-mel spectrogram.

Furthermore, we observe that the excess recovered energy in the generated audio often introduces noise that is easily detectable by human ears. To mitigate this, we impose a penalty on excess energy in the synthetic audio. These two modifications to the multi-scale mel-reconstruction spectral loss result in a more pleasing sound.
\begin{gather}
  \mathcal L_{\mathrm{mel\text{-}energy} }=E_{(s,t,f)} [\left \|\lambda_{e}(\phi_{t,f}^{s} (x)-\phi_{t,f}^{s}(G(x))  )\right \|_{1}  ] \\
  \lambda_{e}=\left\{\begin{matrix}
    2,  & \phi_{t,f}^{s} (x)<\phi_{t,f}^{s}(G(x)) \\
     1, & \phi_{t,f}^{s} (x)>=\phi_{t,f}^{s}(G(x))
    \end{matrix}\right.
\end{gather}

The final perceptual multi-scale mel loss is the sum of the transient loss and the energy loss. 
\begin{equation}
  \mathcal L_{\mathrm{mel} }=\mathcal L_{\mathrm{mel\text{-}transient} }+\mathcal L_{\mathrm{mel\text{-}energy} }
\end{equation}

We use the HingeGAN adversarial loss formulation and the L1 feature matching loss in our model. The loss weights are 20.0 for the improved multi-scale mel loss, 2.0 for the feature matching loss, 1.0 for the adversarial loss and 1.0, 0.25 for the codebook and commitment losses respectively.

%% file: sections/experiments.tex
  \begin{table*}[t]
    \caption{Objective evaluation of LDCodec at 6kbps, compared with Opus, DAC official and FunCodec.}
    \label{tab:different_codecs}
    \begin{center}
      \begin{tabular}{c|c|c|c|c|c}
        Codec         & bitrate  & ViSQOL$\uparrow$ & Mel distance$\downarrow$  & STFT distance$\downarrow$ & Dec GMACs \\ \hline
        Opus          & 12kbps   & 4.11   & 0.910  & 1.342   & -   \\
        Opus          & 16kbps   & 4.22   & 0.766  & 1.202   & -   \\
        Opus          & 20kbps   & 4.26   & 0.674  & 1.097   & -   \\ \hline
        DAC(official)~\cite{kumar2024high}& 6kbps    & 4.17   & 0.799  & 1.337   & 43.3   \\ \hline
        DACLite(Decoder pruning) & 6kbps & 3.97 & 1.055    & 1.514   & 0.28   \\ \hline
        FunCodec~\cite{du2024funcodec} & 6kbps    & 3.89   & 2.052  & 2.351   & $\sim$0.2   \\ \hline
        \textbf{LDCodec} & 6kbps & \textbf{4.14}  & \textbf{0.973}  & \textbf{1.460} & \textbf{0.26} \\ \hline
        
      \end{tabular}

    \end{center}
    \end{table*}
\begin{table*}[t]
  \caption{Ablation studies validated our proposed methods in decoder, quantizer, discriminator and loss function.}
  \label{tab:ablation_exp}
  \begin{center}
    \begin{tabular}{m{2cm}<{\centering}|c|c|c|c|c}
      Ablation on                & model                        & ViSQOL$\uparrow$        & Mel distance$\downarrow$    & STFT distance$\downarrow$   & Dec GMACs \\ \hline
                                 & LDCodec                      & 4.14          & 0.973          & 1.460          & 0.26      \\ \hline
      Decoder                    & $w.o.$ Proposed residual unit     & 3.97          & 1.055          & 1.514          & 0.28      \\ \hline
      \multirow{2}{*}{Quantizer} & $w.o.$ LSRVQ                 & 4.11          & 0.985          & 1.507          & 0.26      \\ \cline{2-6} 
                                 & $w.$ AvgPooling Extractor & 4.11          & 0.991          & 1.513          & 0.26       \\ \hline
      Discriminator              & $w.o.$ Subband-fullband disc & 4.11          & 1.003          & 1.497          & 0.26      \\ \hline
      \multirow{2}{*}{Loss}      & $w.o.$ Transient loss        & 4.09          & 0.999          & 1.497          & 0.26      \\ \cline{2-6} 
                                 & $w.o.$ Energy loss           & 4.15          & 0.943          & 1.486          & 0.26      \\ \hline
        \end{tabular}

  \end{center}
  \end{table*}
\subsection{Datasets and evaluation metrics}

LDCodec was trained on datasets AIshell3~\cite{shi2020aishell} and LibriTTS~\cite{libritts}, all speech was resampled at 16kHz. We adopted the same optimizer configuration as used in DAC~\cite{kumar2024high}. The batchsize and training step were set to 16 and 800k respectively. ViSQOL~\cite{visqol}, Mel distance and STFT distance~\cite{kumar2024high} were adopted to evaluate the objective quality of LDCodec. For subjective tests, out of domain audio samples with a MUSHRA-inspired crowd-sourced method~\cite{mush} were used. 

\subsection{Comparison with other codecs}

In order to assess the speech quality of LDCodec, we evaluated it alongside different codecs using a sample of 40 multilingual speech sequences.
We incorporated the official open-source DAC~\cite{kumar2024high} and the FunCodec~\cite{du2024funcodec} into the evaluation. Moreover, to compare neural audio codecs with similar decoding complexity, we retrained DACLite using our training dataset. DACLite represents the channel-pruning version of the decoder of the official DAC model, and its decoding complexity is 0.28 GMACs.
The objective scores in Table \ref{tab:different_codecs} revealed that the LDCodec achieves comparable ViSQOL scores to the official DAC model, while outperforming DACLite and FunCodec.

\begin{figure}[t]
  \begin{minipage}[b]{1.0\linewidth}
    \centering
    \centerline{\includegraphics[width=1.0\linewidth]{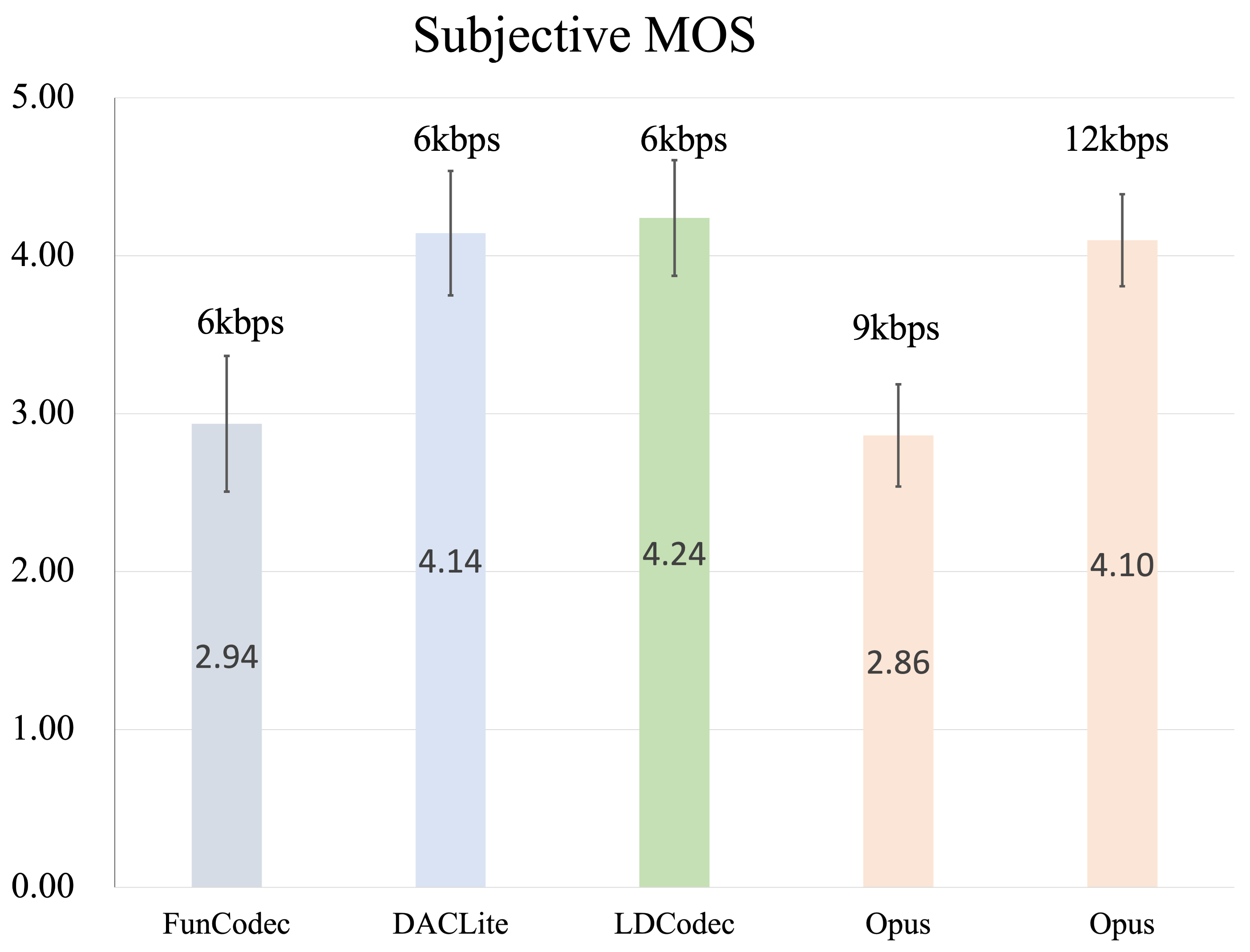}}
  \end{minipage}
  \setlength{\abovecaptionskip}{-0.4cm}
  \caption{Subjective scores for different codecs. Error bars denote the standard deviation.}
  \label{fig:subjective_score}
  
  \end{figure}

As can be seen from the subjective results in Figure \ref{fig:subjective_score}, our proposed codec at 6kbps outperforms Opus\footnote{https://opus-codec.org} at 12kbps.  It's also worth noting that LDCodec at 6kbps excels over FunCodec and DACLite at the same bitrate, which demonstrates the superiority of the LDCodec architecture.

\subsection{Ablation experiments}

A series of ablation experiments were conducted to assess the benefits of incorporating the proposed algorithms into LDCodec. All models operated at 6kbps and the results are shown in Table \ref{tab:ablation_exp}. 

Our findings indicated that our proposed decoder residual unit, which combines lower computational complexity with enhanced coding quality, positively increasing the ViSQOL score from 3.97 to 4.14.
In comparison to factorized rvq in DAC, our proposed LSRVQ leverages inter-frame correlations to significantly enhance the STFT distance from 1.507 to 1.460. 
We also tried LSRVQ with AvgPooling based feature Extractor. However, the experiments indicated a Conv1D based feature Extractor is more efficient.
The introduction of the subband-fullband discriminator and transient loss function further improves objective quality. While the asymmetrical energy loss slightly affects the Mel distance score, it notably enhances the subjective quality because less noisy sound appears in the reconstructed audio.
